\documentclass[fleqn,usenatbib]{mnras}

\usepackage{newtxtext,newtxmath}
\usepackage{multirow,multicol}

\usepackage[T1]{fontenc}

\DeclareRobustCommand{\VAN}[3]{#2}
\let\VANthebibliography\thebibliography
\def\thebibliography{\DeclareRobustCommand{\VAN}[3]{##3}\VANthebibliography}
\usepackage{graphicx}	% Including figure files
\usepackage{amsmath}	% Advanced maths commands

\usepackage{amssymb}	% Extra maths symbols
\title[Multicolor Photometry of MegaCons]{Multicolor photometry of LEO mega-constellations Starlink and OneWeb}

\author[Hui Zhi et al.]{
Hui Zhi,$^{1,2}$\thanks{E-mail: huizhi@nao.cas.cn}
Xiaojun Jiang,$^{1,2}$
Jianfeng Wang,$^{1,2}$\thanks{E-mail: wjf@nao.cas.cn}
\\
$^{1}$CAS Key Laboratory of Optical Astronomy, National Astronomical Observatories, Chinese Academy of Sciences, Beijing 100101, China\\
$^{2}$University of Chinese Academy of Sciences, Beijing 100049, China\\
}

\date{Accepted 2024 February 29. Received 2024 February 21; in original form 2023 November 11}

\pubyear{2024}

\begin{document}
\label{firstpage}
\pagerange{\pageref{firstpage}--\pageref{lastpage}}
\maketitle

\begin{abstract}
  The development of low earth orbit (LEO) mega-constellation fundamentally threatens ground-based optical astronomical observations. To study the photometric properties of the LEO mega-constellations, we used the Xinglong 50 cm telescope to conduct a large-sample, high-precision, and multicolor target-tracking photometry of two typical LEO Mega-constellations: Starlink and OneWeb. Over a three-month observation period starting on 2022 January 1st, we collected 1,447 light curves of 404 satellites in four typical versions: Starlink v1.0, DarkSat, VisorSat, Starlink v1.5, and OneWeb. According to data statistics, Starlink v1.0 has the smallest median magnitude at clear and SDSS $gri$ band, and OneWeb is the dimmest bus. The brightness of Starlink v1.5 is slightly brighter than VisorSat. We construct a detailed photometric model with solar phase angle variations by calculating the illumination-visibility geometry based on the orbital parameters. Our data analysis shows that the solar phase angle is the significant characteristic which influencing Starlink satellites' brightness, but it is not sensitive to OneWeb satellites. VisorSat and Starlink v1.5 version, which are equipped with deployable visors, have significantly reduced scattered light compared to the previous Starlink v1.0 version. The multiband LOWESS and color-index are analyzed in characterizing the energy and color features of LEO mega-constellation satellites. This work found that the proportion of scattered sunlight mitigation achieved with VisorSat and Starlink v1.5 was 55.1\% and 40.4\%, respectively. The color index of different buses shows an evident clustering feature. Our observation and analysis could provide valuable quantitative data and photometric models, which can contribute to assessing the impact of LEO mega-constellations on astronomical observations.
\end{abstract}


%%%%%%%%%%%%%%%%%%%%%%%%%%%%%%%%%%%%%%%%%%%%%%%%%%

%%%%%%%%%%%%%%%%% BODY OF PAPER %%%%%%%%%%%%%%%%%%

\section{Introduction}
\label{sect:intro}
% This is a simple template for authors to write new MNRAS papers.
% See \texttt{mnras\_sample.tex} for a more complex example, and \texttt{mnras\_guide.tex}
% for a full user guide.

% All papers should start with an Introduction section, which sets the work
% in context, cites relevant earlier studies in the field by \citet{Fournier1901},
% and describes the problem the authors aim to solve \citep[e.g.][]{vanDijk1902}.
% Multiple citations can be joined in a simple way like \citet{deLaguarde1903, delaGuarde1904}.

Low Earth Orbit (LEO) refers to an orbit with altitude lower than 2,000 km. Compared with geostationary orbit (GEO), deploying communication satellites in LEO offers advantages such as low launch consumption, low delay, and low transmission loss in satellite communications, but result in shot transit time and poor coverage of a single satellite. Therefore, achieving a global coverage in a LEO communication constellation may require a significant number of satellites. With the development of electric propulsion and microelectronics technology, satellites have gradually developed towards small size, lightweight, low cost, and standardization, which makes building satellite constellations containing thousands of satellites possible.

Several commercial space companies, represented by SpaceX (U.S.) and OneWeb (U.K.), along with several state-owned agencies, have proposed building LEO mega-constellations (MegaCons) comprising thousands of micro and nano communication satellites. These mega-constellations aim to offer reliable, low-latency, high-speed satellite internet services globally. Currently, there are 43 mega-constellation programs still in the development phase. By 2030, the number of operational artificial satellites in near-Earth space is predicted to exceed 100,000 \citep{Krantz+2021}. Some information on these constellations is given in Table.~\ref{Tab:ConsInfo}.

\begin{table*}
	\centering
	\caption[]{Information of Several LEO Mega-constellations}
	\label{Tab:ConsInfo}
	%%Please Capitalize the First Letter of Each Notional Word in table's caption
	
	\begin{tabular}[width = \columnwidth]{cllll}
	  \hline\noalign{\smallskip}
	No. &   Constellation   & Manufacture& Orbit height & Satellites\\
		&	&	&(km)	& No.\\
	  \hline\noalign{\smallskip}
	1&  Starlink Gen1   & SpaceX            	&350-550    &4,408\\
	2&  Starlink Gen2   & SpaceX            	&530-560    &7,500\\
	3&  OneWeb          & OneWeb/Airbus     	&1,200      &900\\
	4&  Project Kuiper  & Amazon            	&590-630    &3,232\\
	5&  Hanwha System   & Hanwha            	&Unknown    &2,000\\
	6&  Globarstar      & Globarstar        	&1,414      &48\\
	6&  Iridium NEXT    & Thalse Alenia Space	&781        &66\\
	%   \noalign{\smallskip}\hline
	\hline
	\end{tabular}
\end{table*}

\begin{figure}
	\centering
	\includegraphics[width = \columnwidth]{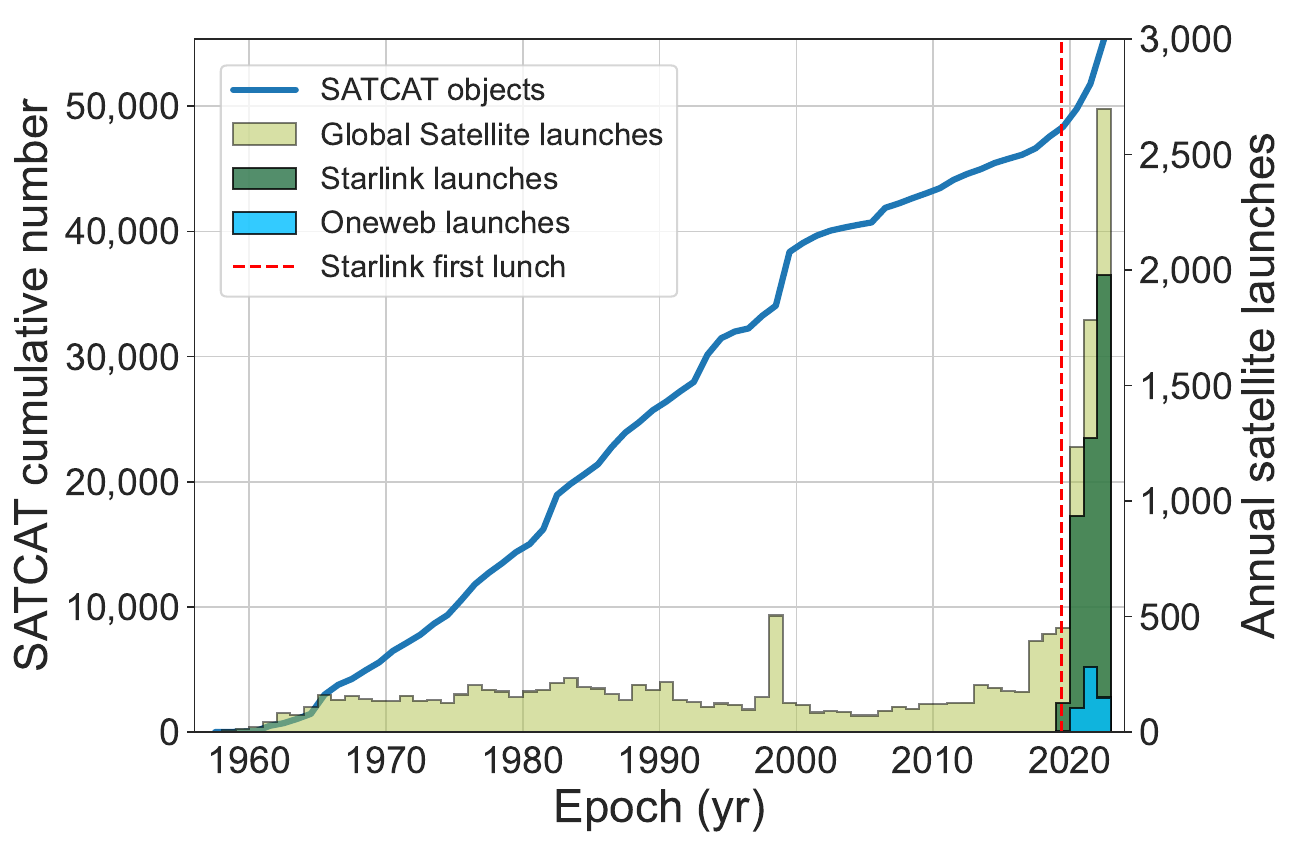}
	\caption{Accumulation of satellites cataloged in the Satellite Catalog (SATCAT). The red dotted line marks the date of Starlink's first launch. }
	\label{fig:satcat}
\end{figure}

\citet{Smith+1982} first addressed that the brightness of satellites, which is primarily derived from scattered sunlight in the optical wavelength range, has a detrimental impact on ground-based astronomical observations. At that time, only approximately 4,600 artificial objects were cataloged in orbit. However, satellite light pollution has progressively increased since the development of LEO mega-constellations. The rapid growth of artificial objects in near-Earth space, as shown in Fig.~\ref{fig:satcat}, poses a significant threat to the safety of space activities and ground-based astronomical observations.

Numerous astronomical studies depend on persistent observations or continuous sequences of stabilized observations, during which the passage of satellites through a telescope's field-of-view (FOV) can lead to stray light interference, detector saturation, flux contamination, and changes in the image background. Such effects have an irreversible impact on astronomical studies such as time-domain astronomy, sky surveys, deep-field observations, spectroscopic observations, and surface source imaging. Wide-field sky surveys, such as the Vera C. Rubin Observatory (VRO, \citet{VRO+2022}), have borne the impact.

On June 29, 2020, the American Astronomical Society hosted the Satellite Constellation 1 Workshop (SATCON1) to assess the impact of LEO mega-constellations on optical observations and propose corresponding mitigation methods. SATCON1 generated several recommendations for satellite operators \citep{Walker+2020}, including the following revisions: 
\begin{enumerate}
	\item Operators should perform adequate laboratory Bi-directional Reflectance Distribution Function (BRDF) measurements as part of their satellite design.
	\item The apparent magnitude of the satellites should be fainter than 7.0 $V_{mag}$, with a normalization distance at 550 km. 
	\item Operators should adjust satellites' attitude and orbit to minimize reflected light and avoid specular reflection on the ground.    
\end{enumerate}

SATCON1 ultimately concluded that no mitigation methods could eliminate the impact of the satellite's trajectory on astronomical observations. According to an agreement reached between NSF and SpaceX in 2022 \citep{NSF+2022}, SpaceX's Starlink constellation and the subsequent second generations (Starlink Gen2) will reduce the orbit altitude no higher than 700 km and brightness of satellites to 7.0 $V_{mag}$ by modifying the satellite design and attitude control strategies. The agreement also stipulated that SpaceX would disclose the orbit information publicly to allow astronomical observation avoidance.

When studying the impact of LEO mega-constellations on astronomical observations, a high-precision brightness model is significant for predicting the variability in their brightness. Due to the stable motion state of GEO satellites, the brightness changes slowly, and a set of more mature observation methods and photometric modeling techniques has been established \citep{Tang+2008}. However, the dynamic nature of LEO satellites poses challenges in their photometric study, including complex variations in brightness, extensive spatial and temporal coverage, and sparse availability of data points. Characterizing the photometric model of LEO satellites has always been a difficult topic. The study of photometric characterization specific to LEO mega-constellations has only taken off in recent years.

\citet{Tyson+2020} investigated strategies for evading the trajectories of LEO mega-constellations on VRO and its LSST telescopes. Observations of selected Starlink satellites in this work were carried out using the Blanco 4 m telescope. \citet{Tregloan+2020,Tregloan+2021} used telescopes to characterize the photometric difference between the low albedo painted STARLINK-1130 (DARKSAT) and the same batch STARLINK-1131 and found that DARKSAT had an average magnitude of $7.46 \pm 0.04$ in the $g'$ band (SDSS photometric system) at a distance of 976.5 km, compared with a magnitude reduction of $0.77 \pm 0.05$ from STARLINK-1131. Mallama has carried out a systematic study of LEO mega-constellations. \citet{Mallama+2021c,Mallama+2022a} used the Russian MMT-9 telescope array, and naked-eye observations for statistically characterized Starlink and OneWeb respectively. \citet{Mallama+2021b} observed 430 Starlink v1.5 version satellites at zenith with an average magnitude of $ 5.92 \pm 0.04 \quad Vmag$, which is 31\% lower than the version without visors. \citet{Mallama+2020} uses the solar phase angle (angle between sun-satellite-station, PSA) as the only parameter of the flat panel model. It fits the Starlink photometric data using a linear or quadratic function to obtain an abbreviated empirical photometric model. A preliminary analysis of Starlink photometry in the terrestrial eclipses was performed in the \citet{Mallama+2021a}. \citet{Boley+2022} utilized the Plaskett 1.8 m telescope to observe 23 Starlink satellites in the $g'$ band. As the telescope could not track LEO satellites, the team extracted brightness from the streaks in 30-second exposure stellar gazing images, and found that the average magnitude at 550 km of visor and visorless satellites was 5.1 and 5.7 $g'_{mag}$. \citet{Krantz+2021} investigated the satellite brightness distribution as a function of the satellite's altazimuth position and time. \citet{Halferty+2022} observed the Starlink constellation using a small aperture, large FOV automated telescope to measure the mean Gaia $G$ magnitude and pointed out that the orbital forecasts of Starlink satellites have errors ranging from 0.12\degr to 0.3\degr in the Dec and R.A. direction, posing a challenge for tracking observations.

Restricted by the observation modes, most of the studies above focus on data statistics or streak photometry, and only short arc segments can be acquired, with insufficient data coverage and accuracy. More high-accuracy observation data and fine photometric modeling for target-tracking observations must be needed. In this paper, through a large-sample multicolor target-tracking observations of two representative LEO mega-constellations, Starlink and OneWeb, we statistically characterize the overall brightness distribution and construct the detailed photometric models based on solar phase angle (SPA). The effectiveness of brightness mitigation measures for different versions of Starlink satellites is investigated through the analysis of observation data. This study provides data support for quantitative analysis of the impact of LEO mega-constellations on optical astronomical observations and offers valuable insights into the LEO satellite photometric characteristic research.

\section{Mega-Constellation and Satellite Design}
\label{sect:bkg}
\subsection{Starlink}
\subsubsection{Constellation design}
\citet{Starlink+2023} is a LEO constellation program proposed by SpaceX, aiming to provide a high-bandwidth, low-latency global satellite Internet connection. Starlink satellites adopt a modular design, which can be quickly manufactured and launched in batches. Starlink utilizes a typical Walker constellation design \citep{Xue+2021}.
As of March 1, 2023, SpaceX has launched 77 batches totaling 4,035 satellites. The Gen1 of the Starlink program included 4408 satellites in five Groups, with the following constellation orbital parameters in Table.~\ref{Tab:orbit}. SpaceX's planned construction of the Gen2 Starlink consists of 29,998 satellites, and the partially FCC-authorized constellation contains 7,500 satellites.

\begin{table*}
\centering
\caption[]{Orbit Specification of Starlink Constellation}\label{Tab:orbit}
%%Please Capitalize the First Letter of Each Notional Word in table's caption

\begin{tabular}{cccccc}
  \hline\noalign{\smallskip}
Group &  Orbit Planes     & Inclination & Height & Satellites per plane & Total Satellites\\
	&  No.	& (\degr) 	& (km) & No.  & No. \\
  \hline\noalign{\smallskip}
Group 1  & 72   &   53.0 &  550 & 22    &   1584    \\ % new variable
Group 2  & 36   &   70.0 &  570 & 20    &   720     \\
Group 3  & 10   &   97.6 &  560 & 43/58 &   508     \\
Group 4  & 72   &   53.2 &  540 & 22    &   1584    \\
Group 5  & 14   &   43.0 &  530 & 43    &   172     \\
  \noalign{\smallskip}\hline
\end{tabular}

\end{table*}

\subsubsection{Satellites Design}
Starlink satellites adopt flat plane design with a body made primarily of aluminum, a bottom surface toward Earth containing high-throughput phased array antennas, a single solar panel on one long stroke. The solar panel has only one degree of rotational freedom, and it adopts the 'shark fin' state as L-shape, which is the typical working condition configuration in the constellation.

Starlink Gen1 has four satellite buses, v0.9, v1.0, and v1.5. A new generation of larger bus, v2.0 (Gen2), is under development. The pathfinder of Gen2, called v2 mini, whose body size is twice as large as that of Gen1, was launched on Feb. 27, 2023, for the first time. Starlink v1.5 is the majority version launched into orbit currently. The bus size is about $1.5 \times 3$ m with a mass of about 227 kg, and the solar panel is about $4 \times 11$ m when unfolded \citep{Gunter+2023}. 
Gen1 Starlink has tried various extinction designs to reduce scattered light pollution, such as using low albedo coating on the bottom of the satellite(DarkSat), adding a deployable visor design (VisorSat), changing the material of the solar panel cells, etc. SpaceX believes that there are better solutions than sunshade and coating due to the difficulty of thermal control and interference with the laser link\citep{Starlink+2022}. Gen2 satellites will replace the cladding material, redirect the solar panels, low albedo components, flight attitude control strategies to reduce brightness.

\subsection{OneWeb}
\subsubsection{Constellation Design}
The \citet{OneWeb+2023} constellation is proposed by a British company OneWeb, formerly WorldVu, to provide global Internet access to individual consumers and airlines. The satellites are manufactured by OneWeb Satellites, a joint venture between Airbus and OneWeb. On February 27, 2019, OneWeb launched its first batch of 6 satellites, and as of March 1, 2023, the company has launched 16 batches of 544 satellites. Currently, 542 satellites are in orbit.
OneWeb's plan consists of 900 satellites, of which 720 are distributed on 18 orbital planes with an inclination of 87.9\degr and an orbital altitude of about 1,200 km. Each orbital plane has 36 working satellites and 4 backups. The satellites are launched multiply to a parking orbit about 450 km, then ascending using an electric propulsion system to a working orbit of 1,200 km. The OneWeb second-generation constellation proposed in 2021 will contain 6,372 satellites.

\subsubsection{Satellite Design}
\begin{figure}
\centering
\includegraphics[width=\columnwidth, angle=0]{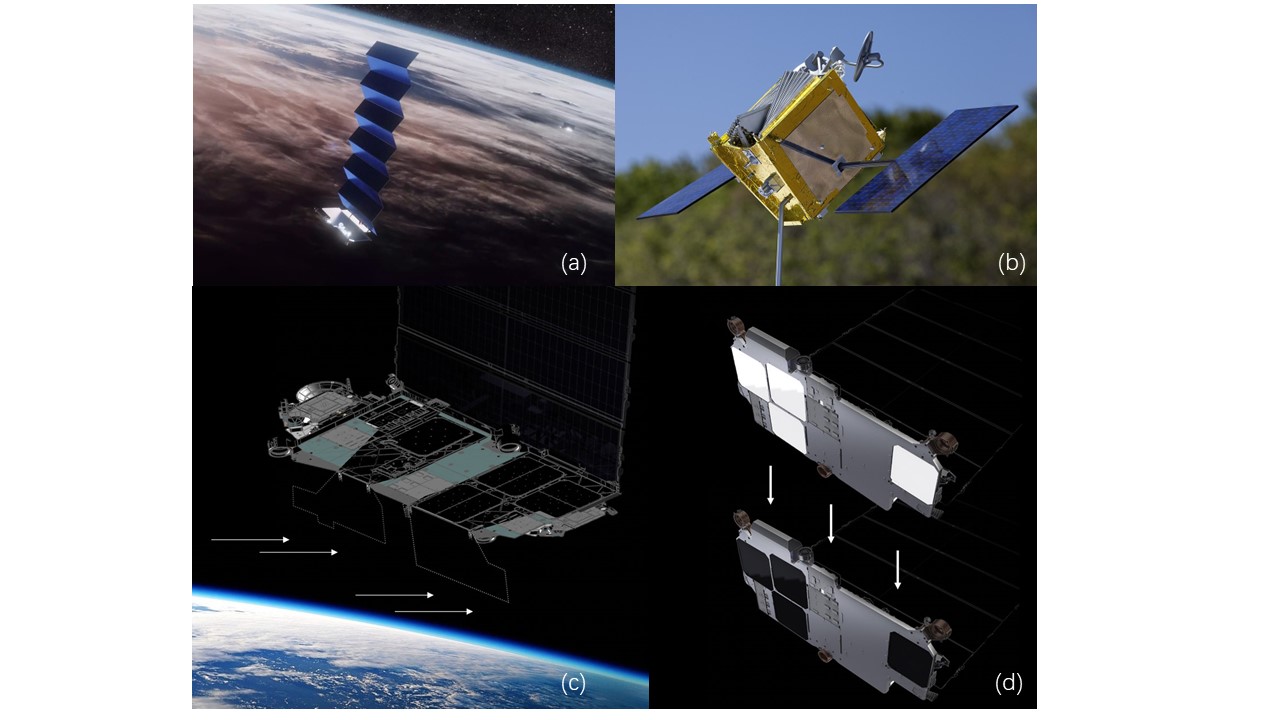}
\caption{Demonstration of 4 satellite buses.  (a) Starlink v0.9 and v1.0 similar (b) OneWeb (c) Starlink v1.0 VisorSat and Starlink v1.5 with deployable visors (d) DARKSAT with a low albedo dark paint design. Source: SpaceX and OneWeb official site.}
\label{Fig2}
\end{figure}

The OneWeb satellite is a 125-150 kg microsatellite with a box design, which is about $1.0 \times 1.0 \times 1.3$ m size. The satellites are designed with two solar panels, and an electric propulsion system for orbital maneuvering, and a service life about five years \citep{Gunter+2023}.
The demonstrated image of Starlink and OneWeb buses involved in this work as shown in Fig.~\ref{Fig2}.

\subsection{The Impact of LEO Mega-Constellations on Astronomical Observations}

The number of artificial satellites in Earth orbit has snowballed since the launch of the LEO mega-constellation. This has resulted in a great concern over satellite light pollution, which can still obscure astronomical observation targets when the satellite is in the Earth shadow, even though its brightness is very low.

\begin{figure}
\centering
\includegraphics[width=\columnwidth, angle=0]{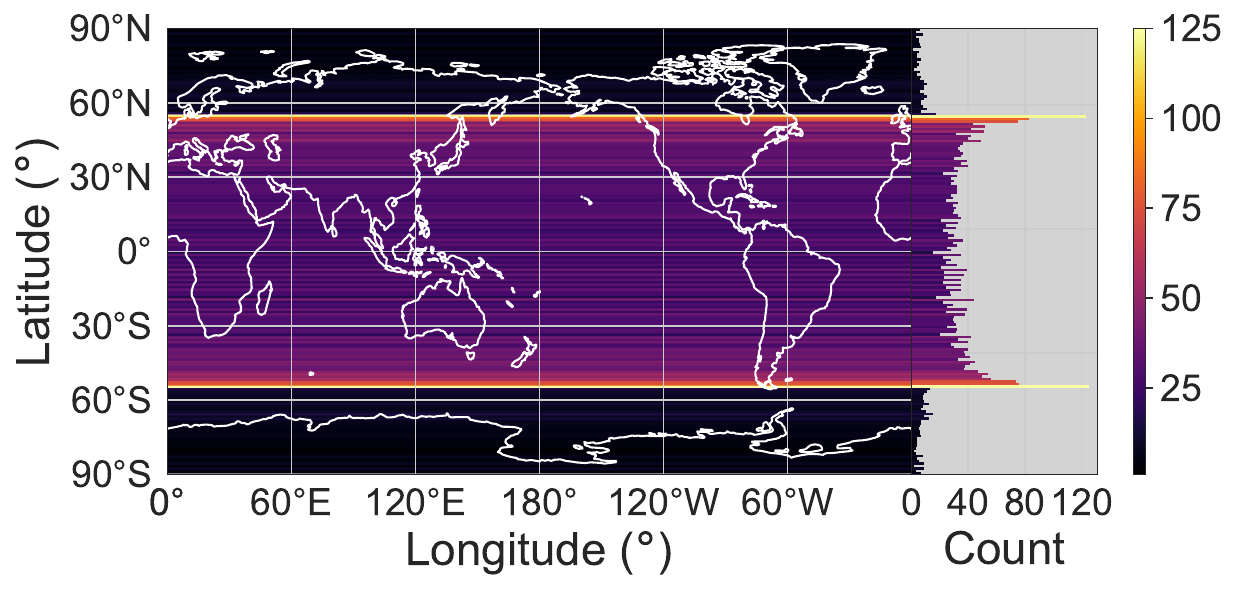}
\caption{ Visible Starlink and OneWeb satellite in different latitude }
\label{fig:subsat}
\end{figure}

\begin{figure}
\centering
\includegraphics[width=\columnwidth, angle=0]{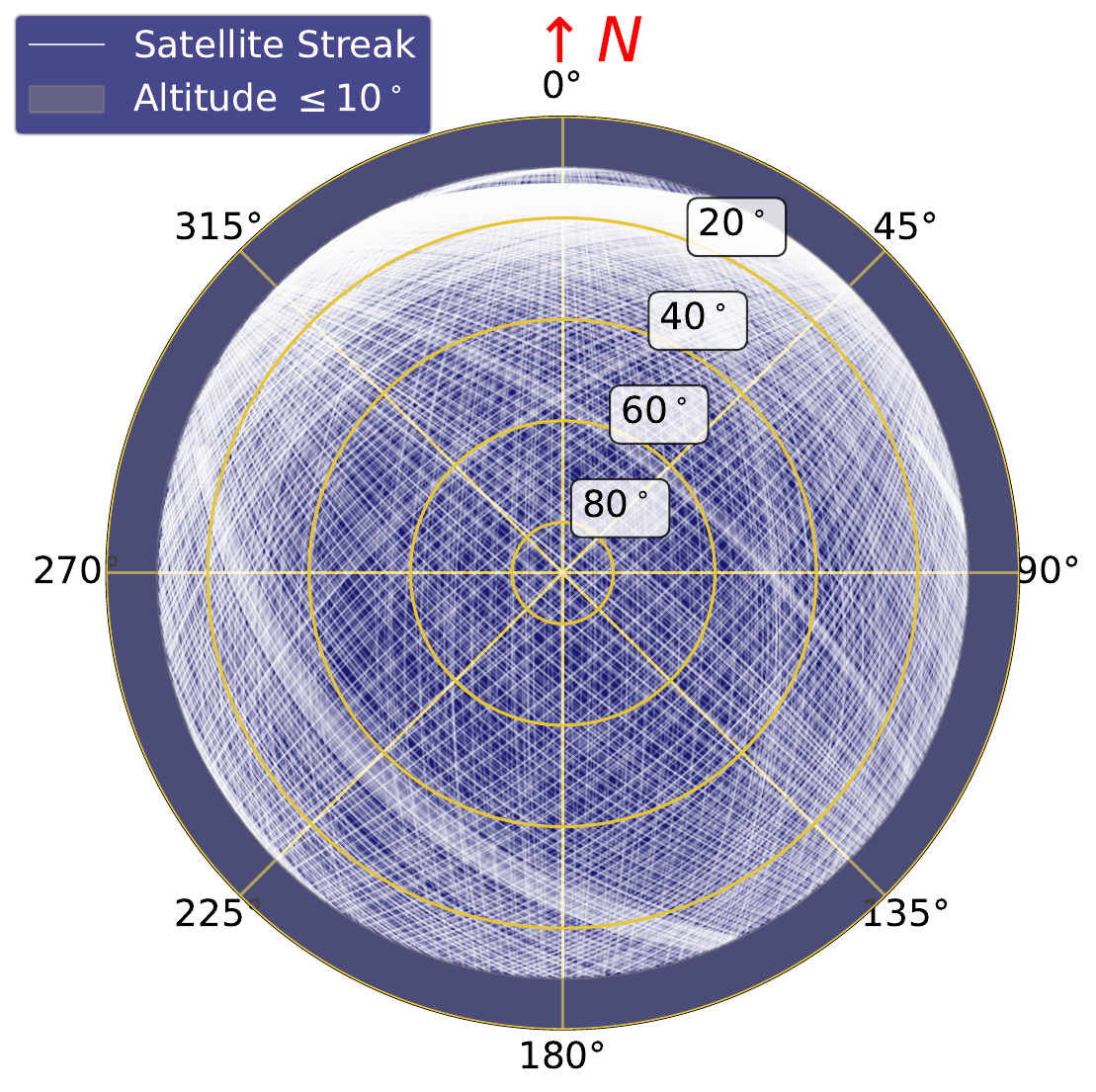}
\caption{Starlink and OneWeb whole night streaks above Xinglong Observatory, 2022 Mar 1st 09:01-23:40 UTC }
\label{fig:arcs}
\end{figure}

In order to quantitatively assess and analyze the impact of LEO mega-constellations on astronomical observations, we calculated the distribution of Starlink and OneWeb constellation sub-satellite position on March 29, 2023, based on public available Two-line element (TLE) files. As depicted in Fig~\ref{fig:subsat}, a high distribution of mega-constellation at mid- to low-latitudes in the northern and southern hemispheres, a range in which many of astronomical observatories around the world are located.
Taking the Xinglong Station of the National Astronomical Observatories, CAS (NAOC) as an example, the constellation transits for the observable period of the night were forecasted based on TLE, with the white line in the Fig.~\ref{fig:arcs} showing the arcs where the transits were forecasted. The coverage was highest at the latitudes of 45-55 \degr north and south hemispheres, resulting in the northern sky of the Xinglong Station being the most seriously affected. As of March 1, the two mega-constellations had a total of 2,359 satellites, and there were 2,683 transits above 10° above the horizon throughout the night, which could influence regular optical astronomical observations.

With the further increase of space objects, the impact on night sky background will no longer be confined to partial regions or streaks but will present a diffused background, resulting in increased sky brightness. Recent studies have shown that this increase will reach about 10\% \citep{Kocifaj+2021}. It is important to note that Starlink satellites cannot actively perceive and avoid debris, so they can only maneuver to avoid collision following ground instructions. In contrast, the vast satellite amount of the mega-constellation may lead to a chain reaction increase in collision debris\citep{Barentine+2023}, which has a higher risk of causing Kessler Syndrome\citep{Kessler+1978}.

\section{Observations}
\label{sect:Obs}
\subsection{Observation mode}
From 2022 January 1, to 2022 March 31, we conducted a large sample of observations of two LEO mega-constellations, Starlink and OneWeb, using the 50 cm telescope at the Xinglong Observatory of the National Astronomical Observatory, Chinese Academy of Science (NAOC). The 50 cm telescope is an equatorial R-C telescope adept at tracking and observing artificial objects, which is capable of simultaneously acquiring the photometric data of the $g$, $r$, and $i$ bands of the SDSS photometric system and can be used for rapid multicolor photometry and analysis of the time domain astronomy, such as transients or variables. Under the target-tracking mode, the overall photometric precision can reach 3.1\% \citep{Jiang+1998,Mao+2012}. 

\begin{figure}
\centering
\includegraphics[width=\columnwidth, angle=0]{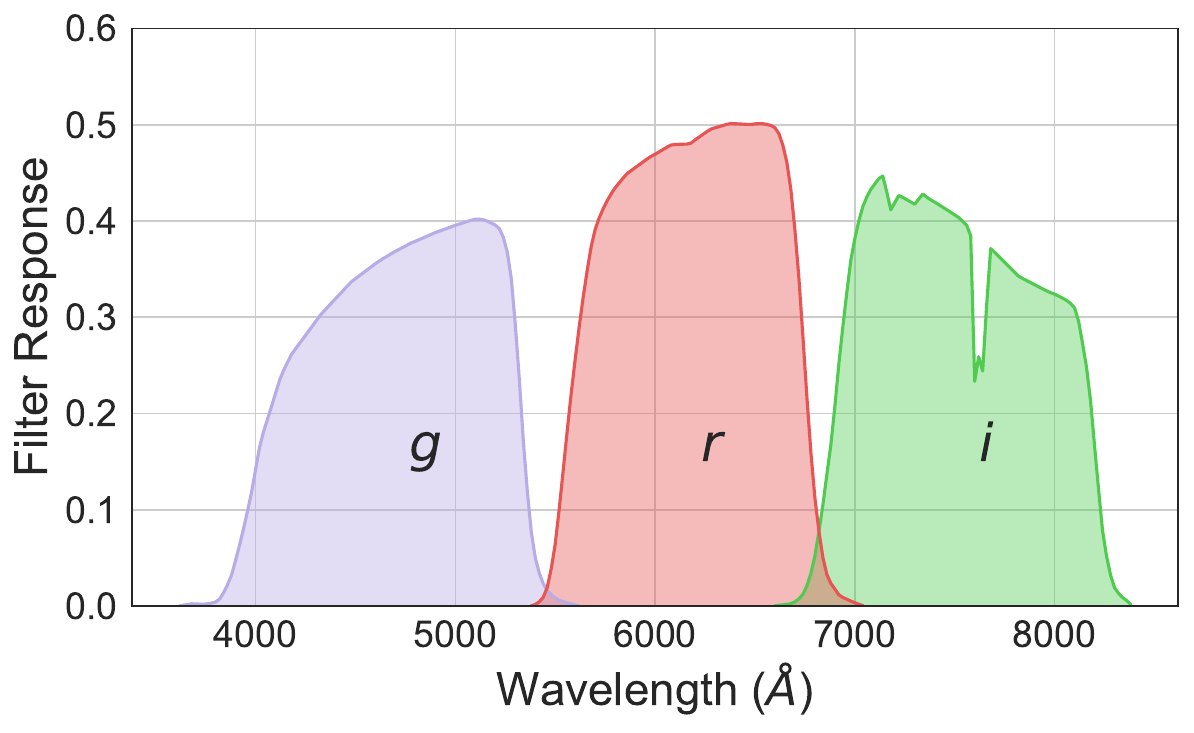}
\caption{50 cm telescope filter passbands curves.}
\label{Fig5}
\end{figure}

The TLEs of mega-constellations were obtained from the CelesTrak \citep{Kelso+2023}. According to the observatory location, we solved a high-precision ephemeris of the satellite's transit arc from TLE, and generated a guidance file for the telescope follow-up satellite's trajectories, called target tracking observation mode. The EMCCDs was used for short-exposure to obtain the high-time resolution data.

\subsection{Data Reduction}
The data processing software developed based on SExtractor \citep{Bertin+1996} to correct the image and extract primary photometry results from images. Information such as time, flux, coordinates, and magnitude of the data points are included.
The magnitude here is the instrumental magnitude, which needs to carry out a series of calibrations and corrections, such as flux calibration, atmospheric correction, slope distance correction, and outlier rejection, to make the satellites' photometry data comparable in different weather, temporal, and spatial environments. The data preprocessing consists of the following methods:

\begin{enumerate}
    \item Flux calibration: Flux calibration converts instrumental fluxes to extra-atmospheric fluxes by observing selected \citet{Landolt+1992} standards stars at different altitudes twice in a photometric night.
    \item Atmospheric extinction reduction: LEO targets have an extensive range of motion in the sky, and the extinction affected by the atmosphere at different orientations in the actual observation is non-negligible. We applied a reduction using Xinglong atmospheric extinction measurement result shown in Table.~\ref{Tab:air_ex}.
    
\begin{table}
\begin{center}
\caption[]{Xinglong Station Atmosphere Extinction Coefficient \citep{Zhao+2020}}\label{Tab:air_ex}
\begin{tabular}{lccc}
  \hline\noalign{\smallskip}
Filter & Extinction Coefficient  & Color Coefficient & RMS\\
	&$k$	& $k_c$&	\\
  \hline\noalign{\smallskip}
g   &$0.269\pm0.019$    &$0.086\pm0.013$    &$0.018$  \\
r   &$0.169\pm0.014$    &$0.125\pm0.009$	  &$0.012$  \\
i   &$0.072\pm0.011$    &$-0.034\pm0.007$	  &$0.008$  \\
  \noalign{\smallskip}\hline
\end{tabular}
\end{center}
\end{table}
    
    \item Slope distance reduction: The slope distance from the target to the telescope varies dynamically from each data point. The scattered light intensity changes with the slope distance according to the inverse square law. Therefore, the concept of absolute magnitude in astronomy is introduced to space object observation. We chose 1,000 km as the standard distance to convert the observed brightness to an absolute magnitude as Equation~\ref{eq1}. 
    \begin{equation}\label{eq1}
        m_d =M-m =  2.5\times \log_{10}{\frac{(1000km)^2}{d^2}}
    \end{equation}
    Where $M$ is absolute magnitude, $m$ is visual magnitude and $d$ is slope distance of data point in kilometers. $m_d$ is the slope reduction coefficient.
    \item Outlier rejection: Influenced by environmental or instrumental factors, there are some outliers in the observational data, which negatively affect the results of data analysis and lead to deterioration of the accuracy of the results. In the data processing, we eliminate them to ensure the accuracy and validity of the light curves. 
\end{enumerate}
According to the above preprocessing method, the conversion relationship between instrument magnitude and calibrated magnitude function shows in Equation~\ref{eq2}:
\begin{equation}\label{eq2}
    m_{cal} = m_{inst} -kA-m_d-Z-k_cC_{std}
\end{equation}
Where $m_{cal}$ is calibrated magnitude, $m_{inst}$ is instrumental magnitude, $C_{std}$ is color index. Airmass $A$ is given as a function of zenith distance $z$\citep{Young+1994}:
\begin{equation}\label{eq3}
    A = \sec{(1-0.0012(\sec ^2 z - 1))}
\end{equation}

The above process generated calibrated photometric results, which are only related to satellites' shape and illumination-visibility geometry, can be compared and analyzed across different datasets.

\subsection{Solve Illumination-Visibility Geometry}
The direct result of data processing is a time-magnitude series (usually refer to light curves) paired with observation time as the only independent variable. However, the light curves at different times may have extensive difference, even for a same target. The brightness of a space object results from the scattered sunlight by its surface materials. It is highly correlated with the observation angle, illumination angle, attitude, and configuration, collectively called illumination-visibility geometry. Therefore, a sequence that describes brightness variation with the satellite's spatial status is needed.

\begin{figure}
\centering
\includegraphics[width=\columnwidth, angle=0]{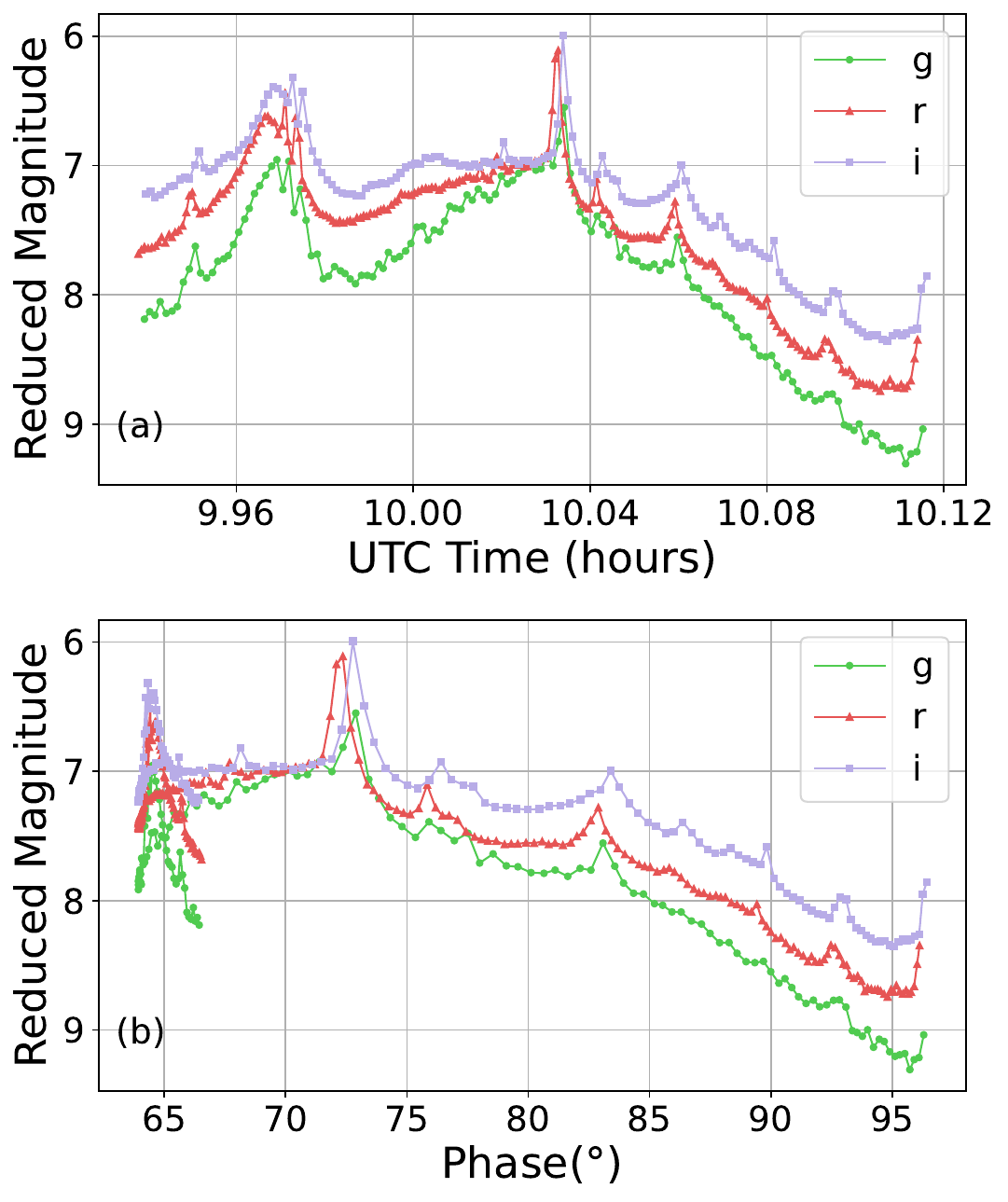}
\caption{Three-channel light curves of OneWeb-0232 (48781). Observed by Xinglong 50 cm telescope. (a) A time-magnitude sequence.(b) A SPA-magnitude sequence solved from TLE }
\label{Fig6}
\end{figure}

Here, we introduce a phase-magnitude series that combines the angles of illumination and observation into a single parameter known as the SPA. An example is shown in Fig.~\ref{Fig6}. Solved from TLE with observation timestamp, the SPA summarized the effects of temporal and spatial variation of a target, allowing for an accurate characterization of the target's illumination and the telescope's visibility. This technique is beneficial for analyzing simple geometrical models, such as flat plates and spherical satellites \citep{Wang+2020}. Overall, the phase-magnitude sequence offers several advantages for space object photometric analysis. 

\section{Data analysis}
\label{sect:analysis}
\subsection{Statistical Analysis}

From 2022 January 1 to 2022 March 31, we observed 329 Starlink satellites with 1051 light curves, and 75 OneWeb satellites with 396 light curves. Starlink's observations covered all 33 launches of in-orbit satellites (as of 2022 February 26). There are four versions of satellites included in the observation: Starlink v1.0, Starlink V1.5, and OneWeb. Starlink v1.0 include the original design, VisorSat and DarkSat. Statistical results of observations are given in Table.~\ref{Tab:stats}.

\begin{table}
\begin{center}
\caption[]{Statistical Results of Observations}\label{Tab:stats}
\begin{tabular}{lccccc}
  \hline\noalign{\smallskip}
Version &\multicolumn{3}{c}{Light curves }  &Total Light Curves   &Total Data Points \\
-   &$g$    &$r$    &$i$    &No.            &No.        \\
    \hline\noalign{\smallskip}
Starlink v1.0   &272    &279    &263    &814    &24276  \\
Starlink v1.5   &78     &73     &32     &223    &9810   \\
Falcon 9 Deb    &9      &7      &8      &24     &467    \\
OneWeb          &134    &141    &140    &415    &36277  \\
  \noalign{\smallskip}\hline
\end{tabular}
\end{center}
\end{table}

Due to the lack of information disclosure, it is impossible to confirm whether VisorSat and Starlink v1.5 had their visors deployed or even equipped with a visor. According to \citet{SpaceX+2020}, satellites before the 10th launch  (L9) did not have a visor, and Starlink-1436 was the first visor experimental satellite. We postulated that satellites launched after Starlink-1522 have deployable visors, i.e., all 57 Starlink satellites launched on 2020 August 6 are the first batch of VisorSat, and this work uses this as a watershed for distinguishing between Starlink v1.0 satellites. Observations of SpaceX's experimental low-albedo coated satellite, STARLINK-1130 (DARKSAT), were also made to compare the photometry characteristic of different light mitigation designs. The observed launches' coverage is shown as Fig.~\ref{Fig6} below:
\begin{figure}
\centering
\includegraphics[width=\columnwidth, angle=0]{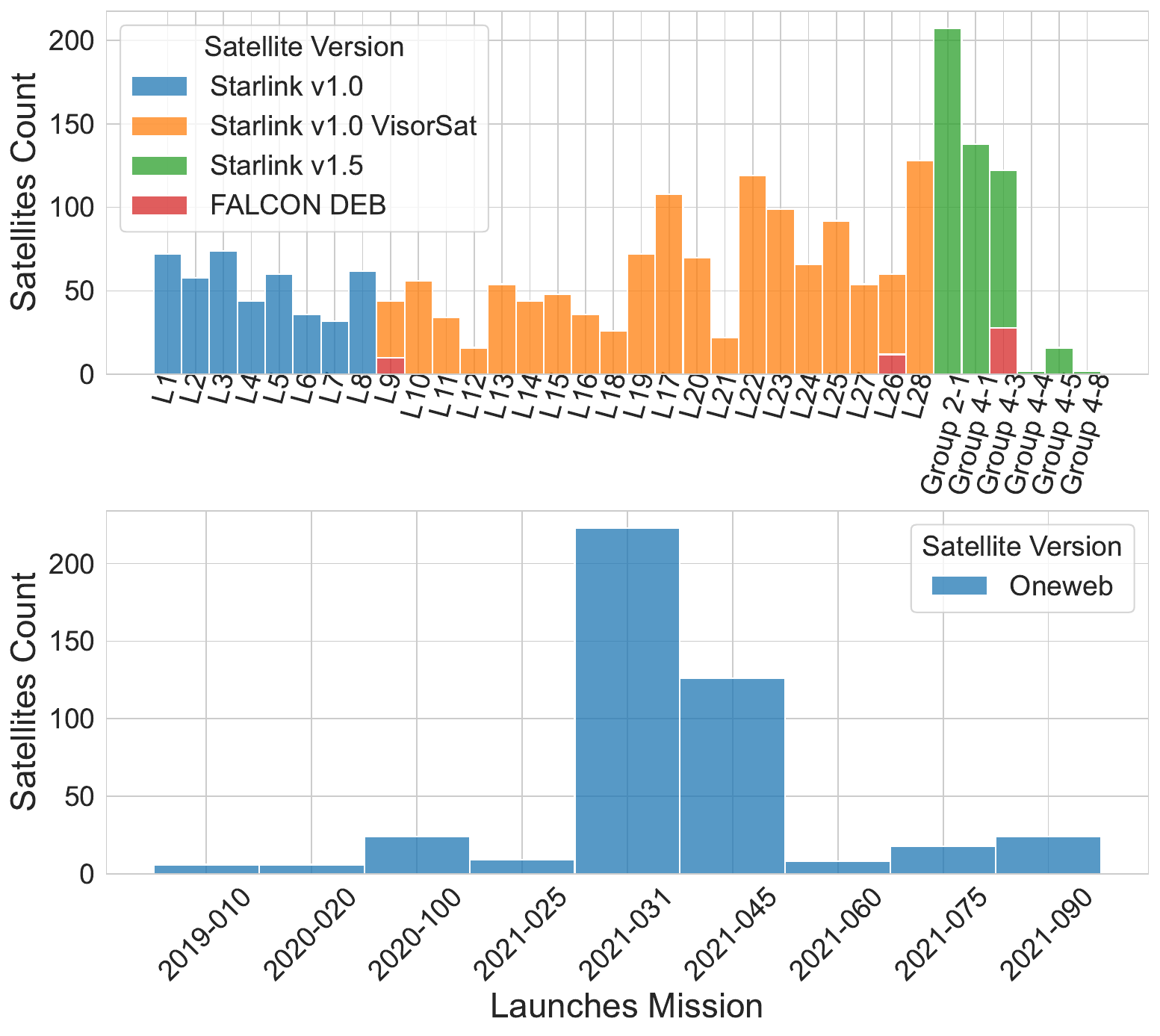}
\caption{Observed launches of Starlink and OneWeb }
\label{Fig7}
\end{figure}

Furthermore, we chose open-source data captured by the MMT-9 telescope array \citep{Karpov+2016}, which observed the same target set for a comparative analysis. The unfiltered data from MMT-9, labeled as $Clear$, is similar to the Johnson $V$ band and has been calibrated to a slope distance of 1,000 km. The statistical distribution of absolute magnitude under all SPA range is illustrated in Fig.~\ref{Fig8} and Table.~\ref{Tab:bar}, where $M$ represents median magnitude and $\sigma$ represents standard deviation:

   \begin{figure}
   \centering
   \includegraphics[width=\columnwidth, angle=0]{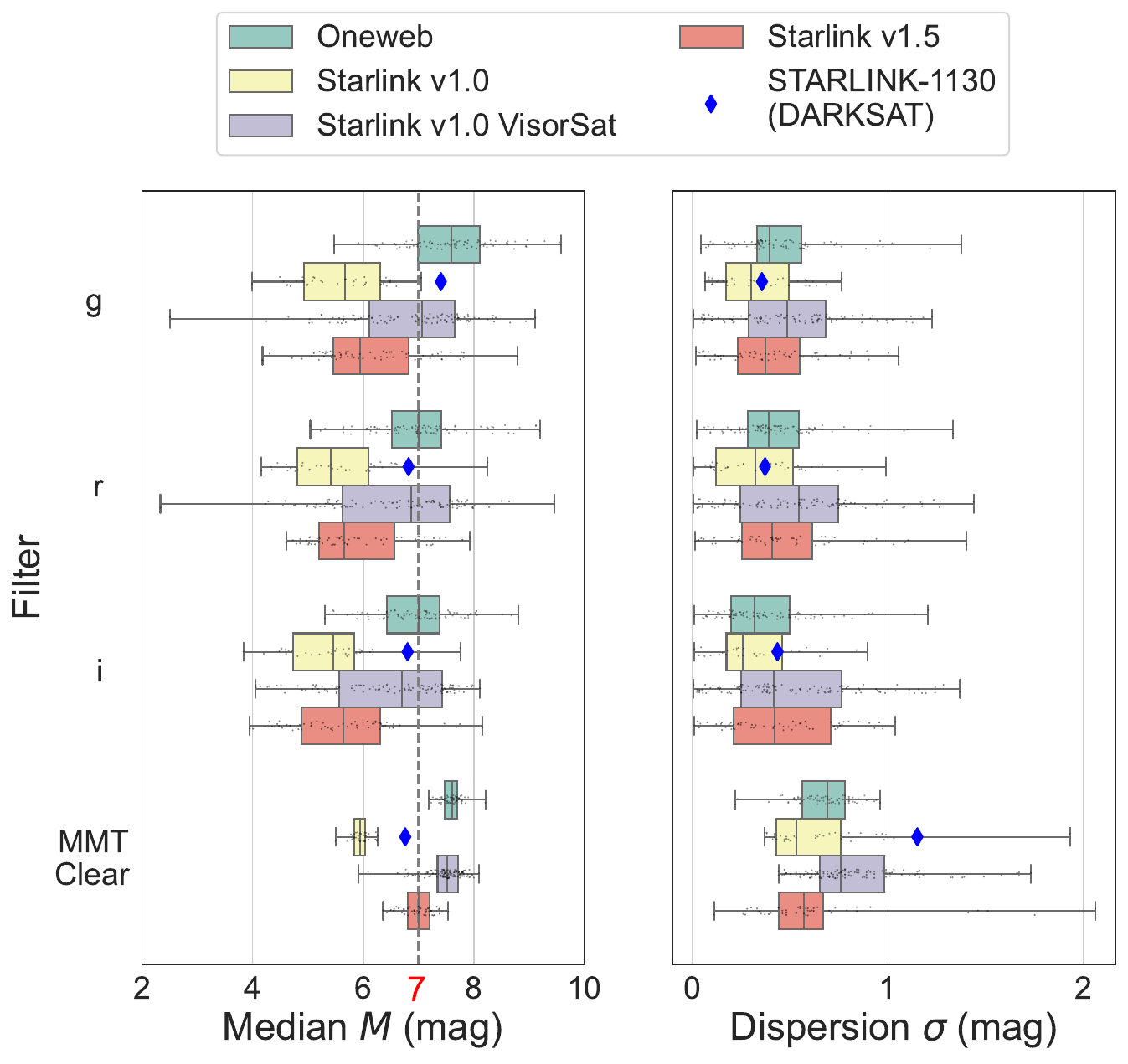}
   \caption{The absolute magnitude distribution of four mega-constellation Bus and DARKSAT }
   \label{Fig8}
   \end{figure}

Based on a statistical analysis of satellite platforms, it has been found that the photometric distribution of Starlink v1.5 satellites is notably brighter than VisorSats. Additionally, there is a significant increase in photometric data dispersion in Starlink satellites equipped with sun visors. Despite ongoing launches, it appears that latest version of Starlink has yet to meet the SATCON agreement \citep{SpaceX+2020}.
Besides, under certain phase angles(sharp angles), the brightness of a flat panel satellite, such as Starlink v1.0, primarily reflects the contribution of the maximum surface of the satellite bodies. The satellite attitude has a minor influence and shows a smaller dispersion. Conversely, cube satellites exhibit a more complicated scattering geometry relationship with attitude variation, providing more mirror reflection opportunities. The phase-magnitude diagram further corroborates this.

\begin{table}
\begin{center}
\caption[]{Starlink and OneWeb Absolute Magnitude Statistics}\label{Tab:bar}
\begin{tabular}{lcccccccc}
  \hline\noalign{\smallskip}
Filter &\multicolumn{2}{c}{OneWeb}   &\multicolumn{2}{c}{Starlink v1.0} &\multicolumn{2}{c}{VisorSat}    &\multicolumn{2}{c}{Starlink v1.5}  \\
-   &$M$    &$\sigma$    &$M$    &$\sigma$    &$M$    &$\sigma$    &$M$    &$\sigma$        \\
- &\multicolumn{8}{c}{(mag)}   \\
    \hline\noalign{\smallskip}
Clear &7.58 &0.68 &5.97	&0.62	&7.52	&0.82	&7.11	&0.68   \\
g	    &7.64	&0.59	&5.85	&0.45	&7.10	&0.57	&6.41	&0.56   \\
r	    &7.08	&0.53	&5.71	&0.44	&6.96	&0.61	&6.32	&0.61   \\
i	    &6.97	&0.49	&5.67	&0.33	&6.83	&0.54	&6.00	&0.53   \\

  \noalign{\smallskip}\hline
\end{tabular}
\end{center}
\end{table}

It is essential to recognize that the photometric result of LEO satellites may be inconsistent and subject to fluctuations based on illumination-visibility geometry variations. Simply examining data distribution from a single position (such as zenith) or using statistical full-arc magnitudes may not fully capture the photometric model. Therefore, We utilize SPA as an effective characteristic for analyzing photometric variations to gain a more accurate understanding.

\subsection{Solar Phase Angle Function Fitting}

\begin{figure}
\centering
\includegraphics[width=\columnwidth]{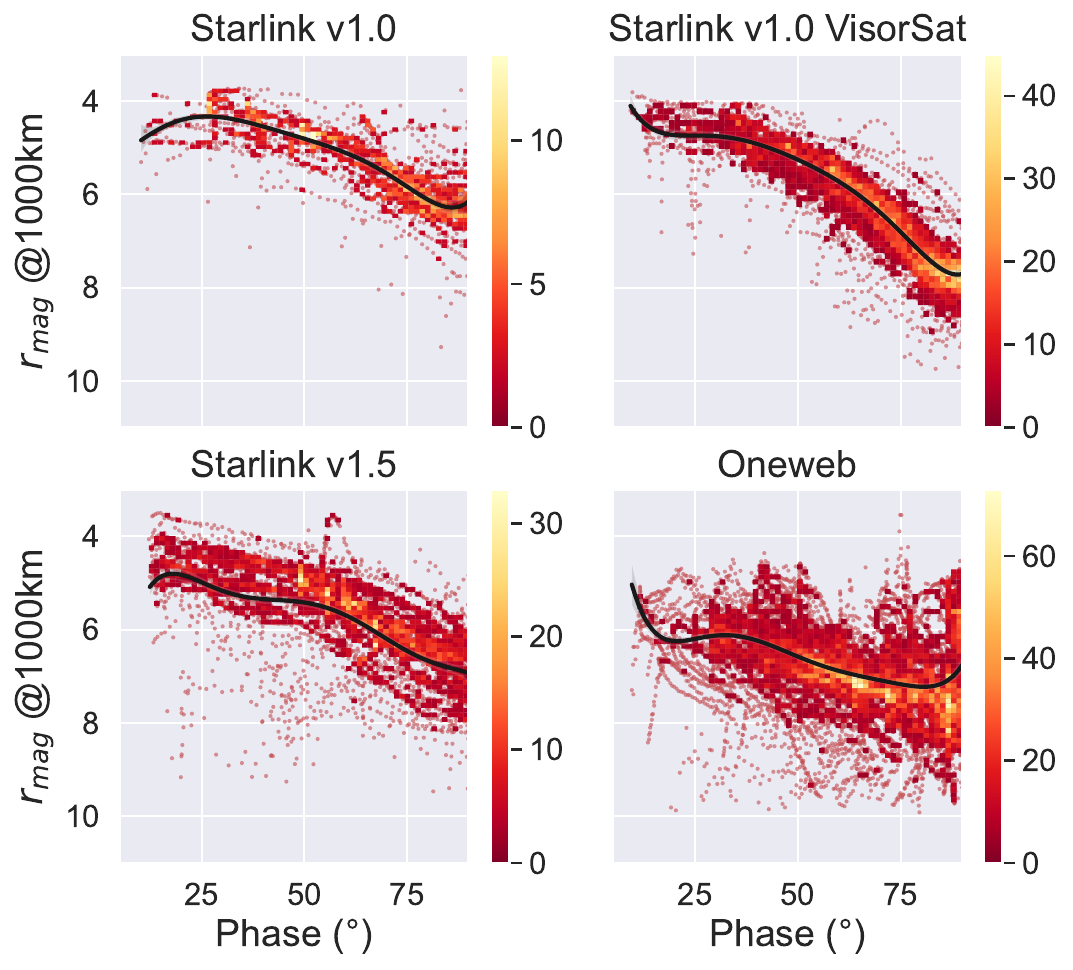}
\caption{\label{fig:rband_fit}{\small Polynomial fitting results of $r$-band SPA-magnitude function} }
\end{figure}

The SPA is crucial for comparing observation data of the mega-constellation across different orbits. To conduct comparison, light curves are folded and aligned to the SPA range. The observation data SPA coverage ranges from  5\degr to 150\degr. Due to the optical observation usually conduct at larger solar enlongation, LEO satellites at small or medium SPA have a greater impact on astronomy. Thus, we focus an SPA smaller than 90\degr for phase analysis in this work. Within this range, we can disregard the contributions of the earth illumination and solar panel backside brightness.
Referring to the empirical formula for the albedo of non-Lambertian spherical satellites proposed by \citet{Cognion+2018}, a sixth-order polynomial is used to fit the SPA-magnitude curve when the SPA is under than 90\degr. SPA-magnitude function fitting result with the goodness-of-fit value, $R^2$, are as shown in Fig.~\ref{fig:rband_fit} and Table.~\ref{Tab:R_suqare} respectively:

\begin{table}
\begin{center}
\caption[]{Goodness-of-fit of Mega-constellations SPA-magnitude Function ($r$-band)}\label{Tab:R_suqare}
\begin{tabular}{lc}
  \hline\noalign{\smallskip}
Satellites		&$R^2$  \\
    \hline\noalign{\smallskip}
Starlink v1.0   &0.67   \\
VisorSat		&0.68   \\
Starlink v1.5   &0.40   \\
OneWeb          &0.16   \\
  \noalign{\smallskip}\hline
\end{tabular}
\end{center}
\end{table}

% \begin{figure}
% \centering
% \includegraphics[width=\columnwidth]{pics_rev/fit_compare.pdf}
% \caption{\label{fig:fit_compare}{\small $r$-band SPA-magnitude Function Comparison}}
% \end{figure}

\begin{figure}
\centering
\includegraphics[width=\columnwidth]{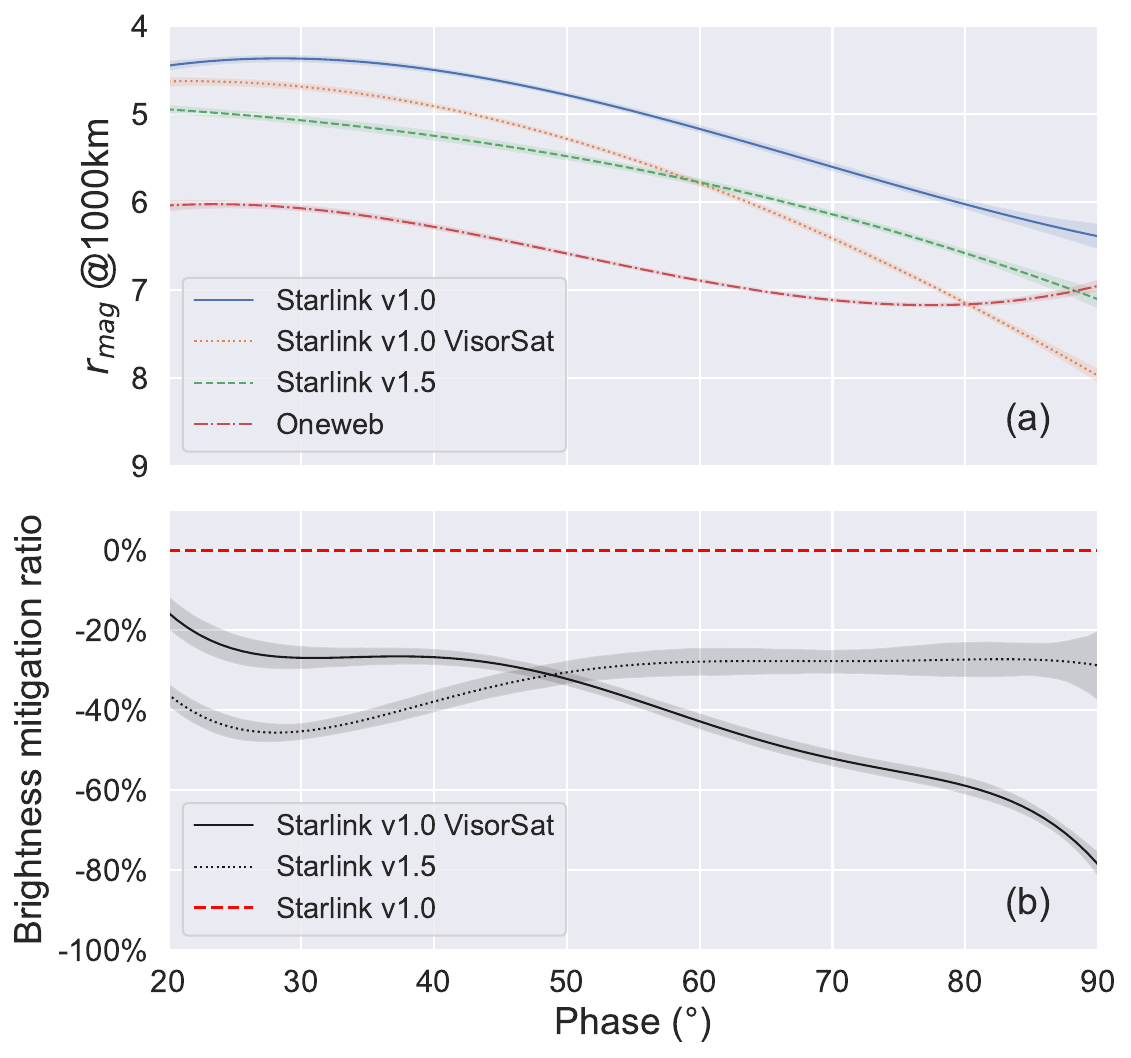}
\caption{\label{fig:fit_compare}{\small (a). $r$-band SPA-magnitude function comparison; (b). Brightness mitigation ratio of VsiorSat and Starlink v1.5 satellites relative to Starlink v1.0}}
\end{figure}

Upon analyzing the fitting results, it was discovered that the OneWeb satellites have fainter brightness than the Starlink where SPA smaller than 80\degr. The reason behind this is the small cross-sectional area of the satellite body. As the SPA approaches 90\degr, the illuminated area of the flat plate satellites reaches a minimum, causing them to be fainter than the OneWeb satellites.
The SPA-magnitude curves of the Starlink versions exhibit significant differences. The Starlink v1.0, which lacks extinction measures, generally has brighter SPA-magnitude curves in the range of 10-90\degr. VisorSat, on the other hand, has a maximum luminosity at around 30\degr, and its brightness decreases significantly with the SPA and is darker at the small SPA. The overall brightness of Starlink v1.5 is fainter than v1.0 satellites, but it is brighter than VisorSat at large SPA.

\begin{figure}
    \centering
    \includegraphics[width=\columnwidth]{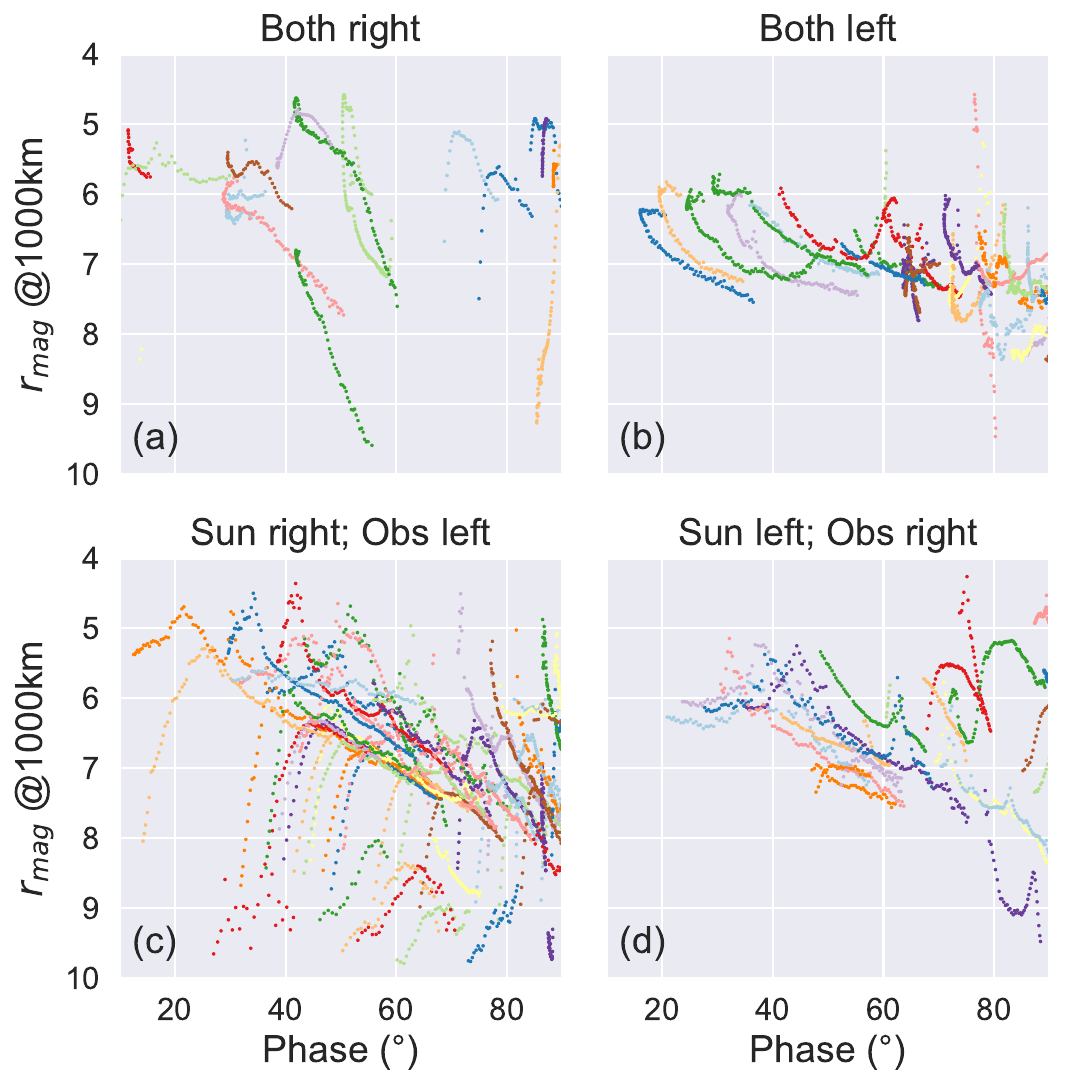}
    \caption{\label{fig:sat_ori}{\small OneWeb's photometry data categorized by the orientation of sunlight(Sun) and observer(Obs); each color represents a light curve (a). The Sun and the observer are on the right side of the orbital velocity direction; (b). The Sun and the observer are on the left side; (c). The Sun is on the right side of the orbit plane while the observer is on the left side; (d). The symmetrical situation of sub-figure (c).}}
\end{figure}

Using solar phase angle as a simplified model to predict magnitude offers computational advantages. However, the OneWeb data shows that the phase curves are less regular with variable turning points or folds.  To improve the precision of the phase model, additional details such as the orientation of the orbit plane should be incorporated.
Utilizing the computation of TLE, we conducted a preliminary analysis of scenarios where the Sun and the observer are on either the same or opposite sides of the OneWeb satellite. The data was categorized into three distinct situations and shown in Fig.~\ref{fig:sat_ori}:

\begin{enumerate}
  \item When both the Sun and the observer on the right side of the orbital velocity direction;
  \item When they are on the left side;
  \item When the Sun and observer are positioned on both sides, containing two symmetrical situations.
\end{enumerate}

According to the phase-magnitude curves, we found that when the Sun and the observer are on the same side, the minimum phase angle is the maximum brightness of the curve, which agrees with the traditional empirical models of GEO or HEO satellites. Nevertheless, when the Sun and observer are not on the same side, the magnitude decreases dramatically for small phase angles instead of peaking earlier. 
Furthermore, the light curve exhibits inconsistencies for instances involving symmetrical illumination—such as when the station and Sun are on the same side. These inconsistencies may be attributed to the inherent asymmetry of the satellite or view position.

\subsection{Brightness Mitigation Assessment}
Since the Starlink Gen1 satellites are all based on the Starlink v1.0 platform, we took the fitted SPA-magnitude function of the Starlink v1.0 satellites as the reference, calculated the differential photometry of the remaining two versions of the satellites at each phase angle, and extrapolate to the scattered light mitigation ratio according to the following equation:

\begin{equation}\label{eq4}
    dm(\phi) = m(\phi) - m_{r}(\phi) = -2.5\lg \frac{F(\phi)}{F_{r}(\phi)}
\end{equation}

\begin{equation}\label{eq5}
   R(\phi) = \frac{F(\phi)-F_r(\phi)}{F_r(\phi)} = (\frac{1}{2.5^{dm(\phi)}}-1)\times 100\%
\end{equation}

Where $\phi$ represents the SPA. $F(\phi),m(\phi)$ represent the specified  bus flux and magnitude function with the variation of $\phi$, and $m_r(\phi),F_r(\phi)$ are Starlink v1.0 fitted magnitudes and flux function in the same SPA range. As shown in Equation~\ref{eq4}, the differential brightness of specified bus and Starlink v1.0 gives as $dm(\phi)$. The mitigation proportion function, $R(\phi)$, is deduced by Equation~\ref{eq5}, and then the following mitigation curves and their confidence intervals are obtained by fitting the data.

Due to the difference in data distribution, the fitting function of Starlink v1.0 may not convincible at minimal SPA (below 10\degr). While this SPA range is not discussed in detail at present, we can report that after performing statistical full-SPA data filtering based on 3-$\sigma$ cropping, the median mitigation ratio of VisorSat and Starlink v1.5 were obtained as 55.1\% and 40.4\%, respectively. VisorSat's brightness mitigation ability is superior under minor SPA conditions.
When examining the mitigation ability of satellites at different SPA, it was observed that at an SPA of 30\degr, the VisorSat's brightness was reduced by approximately 25\% compared to its original version. In contrast, the Starlink v1.5 version was reduced by nearly 50\%. However, at around 46\degr, the brightness of both satellites flips, resulting in better extinction of the VisorSat. This mitigation ratio averages up to 75\% at a phase angle of 90\degr due to the visor that reduces the satellite's scattered sunlight efficiently. On the other hand, the Starlink v1.5 version has an extinction ratio of approximately 25\%, with significant reductions as the phase angle approaches vertical.

\subsection{Multicolor Feature}

When conducting multicolor photometric analysis, the data sampling rate of each band can significantly impact the fitting of features, resulting in notable differences in multicolor photometric trends. In our analysis, we observed that the SPA-magnitude distributions of the LEO mega-constellations had a high degree of dispersion, and the commonly used 6th-order polynomial method for constructing the GEO phase function did not yield optimal results. 

\begin{figure}
  \centering
  \includegraphics[width=\columnwidth, angle=0]{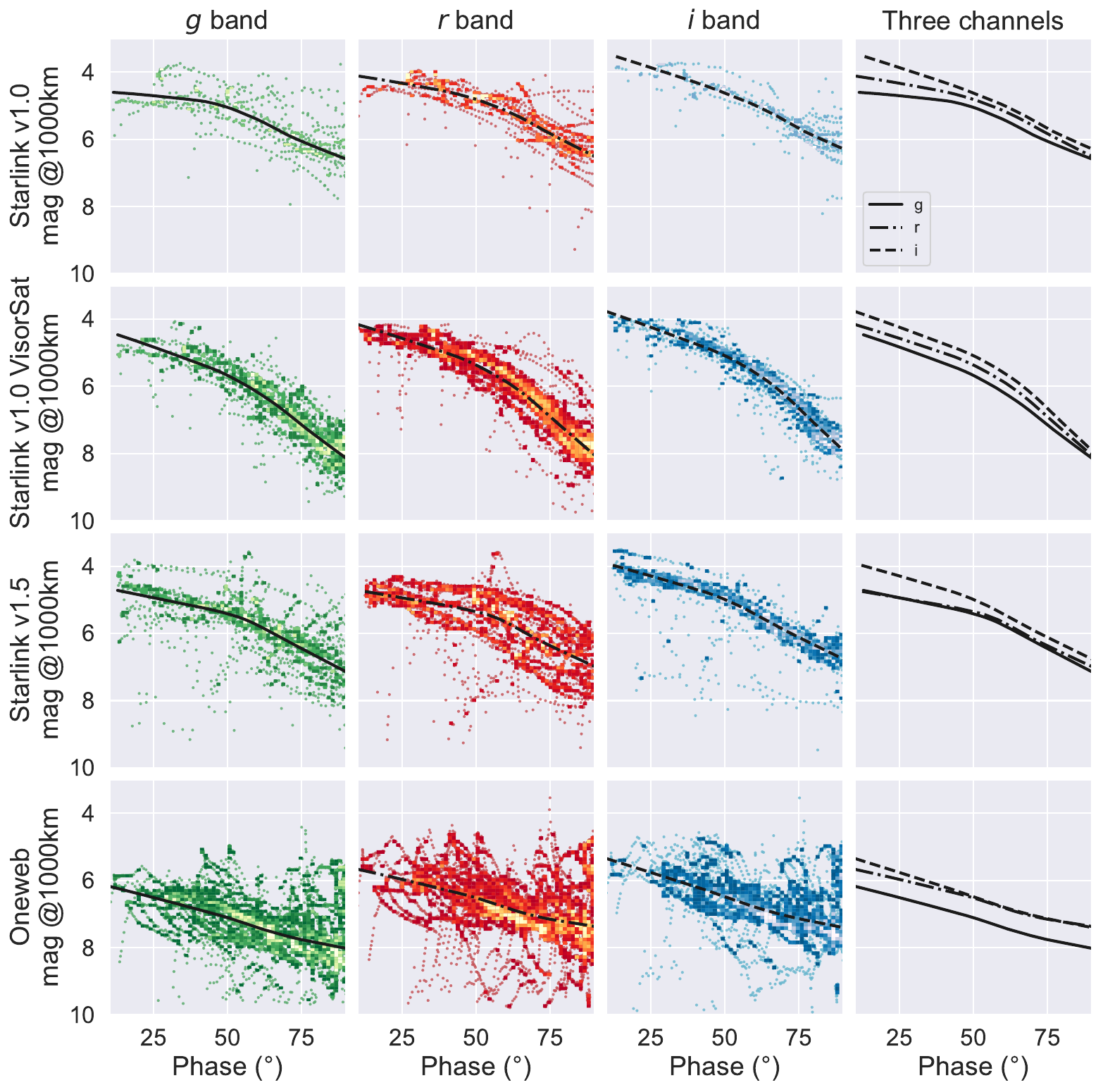}
  \caption{LOWESS analysis of multicolor photometry for Starlink and OneWeb satellites}
  \label{Fig:multi_color_lowess_dark}
\end{figure}

In light of this, we turned to the Local Weighted Regression Scatter Smoothing (LOWESS) algorithm to conduct a regression analysis of the scatter points. This non-parametric regression method is advantageous as it can dynamically fit curve conclusions based on the data near each point without assuming any particular data distribution. Additionally, LOWESS can adapt to nonlinear data and has a robust immunity to outliers and noise. We conducted regression analysis on the multicolor observations from each satellite bus, and the results are depicted in Fig.~\ref{Fig:multi_color_lowess_dark}.

The regression result shows that Starlink and OneWeb satellites exhibit smaller red-end magnitude, which is consistent with the scattering of sunlight. Additionally, the multicolor brightness of both Starlink v1.0 and VisorSat display a similar pattern, with the $i$-band being the brightest, followed by the $g$-band, and the $r$-band being the darkest. Notably, the multicolor photometric distributions of Starlink v1.5 and OneWeb satellites demonstrate considerable variations in phase angle, which may be linked to differences in the scattering material.

\begin{figure*}
\centering
\includegraphics[width=2\columnwidth,angle=0]{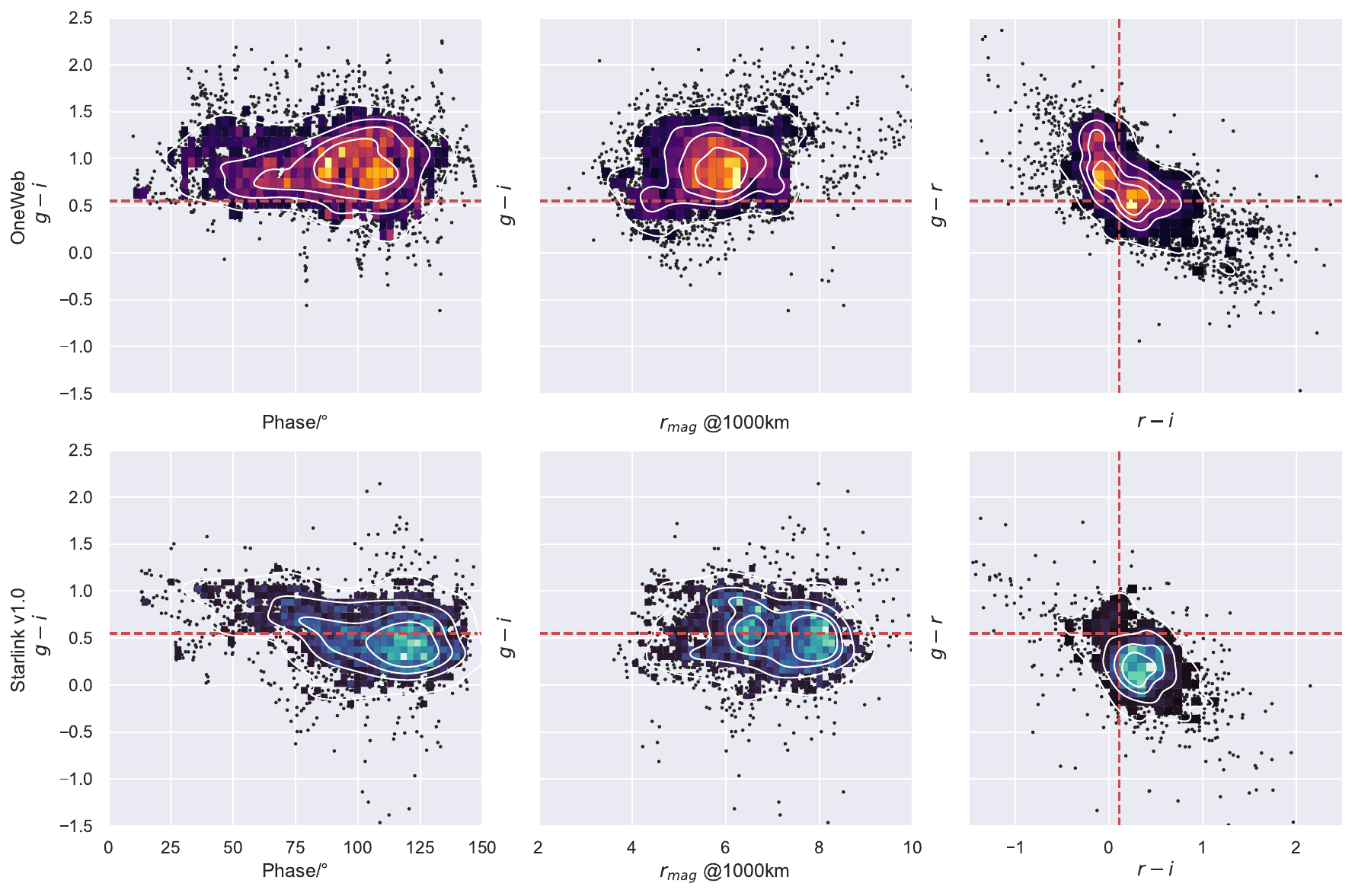}
\caption{Color index diagram of OneWeb and Starlink v1.0 satellites}
\label{Fig13}
\end{figure*}

In addition to multiband light curves, the color index can provide valuable information about the color characteristics and energy distribution of different satellites. This distribution reflects the ability of satellites to absorb and reflect sunlight, thus characterizing their surface material features. Here, we calculated the color index under the same observation conditions through phase synchronization and time synchronization. A total of 11900 valid data pairs of color index were obtained, comprising $g-r$, $r-i$, and $g-i $ color index. The median data synchronization accuracy was 0.0006\degr, and the average time accuracy was 0.86 seconds. 
We selected representative OneWeb and Starlink v1.0 satellites to explore the color differences between them and conduct a kernel density estimation (KDE), as shown in Fig~\ref{Fig13}. The phase and $g-i$ color index diagram in the first column of the figure represents the color characteristics of the satellite at different SPA. The magnitude-$g-i$ color index diagram in the second column reflects the color characteristics of the satellite at different brightness levels. In contrast, the $g-r$ and $r-i$ double color-index diagram in the third column reflects the color tendency.

Since the solar color index of the 50cm telescope's passband \citep{SDSS+dr16+2018} shows in Equation~\ref{eq6}, we can find that OneWeb and Starlink show a distinguished color feature.
\begin{equation}\label{eq6}
  \begin{aligned}
    g-r &= 0.44\pm 0.02 \\
    r-i &= 0.11\pm 0.02 \\ 
  \end{aligned}
\end{equation}

\begin{figure}
  \centering
  \includegraphics[width=\columnwidth, angle=0]{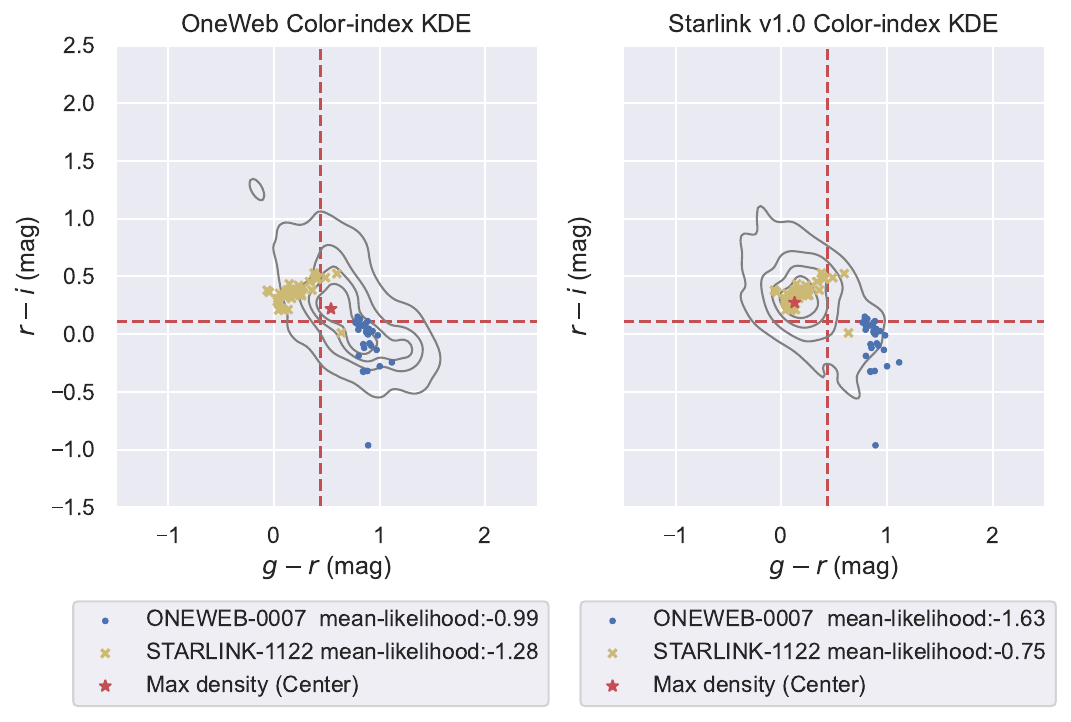}
  \caption{Color index identification of two light curves OneWeb-0007 and Starlink-1122}
  \label{Fig:kde}
  \end{figure}

The observation of the color index shows clustering distribution, indicating similar bus share designs share similar color features. However, the color index varies significantly with brightness or phase angle because a single illumination relationship may correspond to multiple cross-sections and materials of satellites due to changes in attitude. Therefore, a more detailed color index necessitates analysis from more comprehensive satellite orbits, illuminations, and attitude angles. As a result, identifying targets through a single or few data points becomes challenging. 
Nevertheless, color index can still serve as an effective auxiliary method. Two color index curves of OneWeb-0007 and Starlink-1122 satellites are shown in Fig.~\ref{Fig:kde}. The gray contours in the background are the kernel density estimation(KDE) results of the double-color-index distributions of OneWeb and Starlink v1.0 buses. We statistically compared the average likelihoods across two data groups under the observational KDE of the color index. The results indicate that the clustering distribution of the color index is feasible for identifying long-term observation curves of the color index.

In general, the color of OneWeb satellites is redder than sunlight under full SPA and brightness conditions, while Starlink is closer to sunlight. The color index of Starlink varies with the SPA, indicating that its visible and illuminated cross-sectional ratio has changed, while OneWeb's is relatively stable in this situation. Compared with the magnitude-color-index diagram and SPA-color-index diagram, the clustering effect in the double-color-index diagram is almost unaffected by the SPA(except for a small phase angle near specular reflection), with Starlink showing more obvious clustering and OneWeb presenting a dual-kernel characteristic. Based on the above demonstration, the color index of an observation sequence—for instance, multiple measurements and variation of a specific target can be used effectively for LEO constellation satellite identification and cataloging.

\section{Discussion and Conclusions}
\label{sect:discussion}
The present study entails an extensive analysis of two representative LEO mega-constellations, Starlink and OneWeb, based on vast sample observations.
Various data statistical results, phase function characteristics, as well as multicolor features were investigated to examine the photometric properties of the mega-constellations. In light of the findings, the conclusions drawn from this analysis are presented below:
\begin{enumerate}
  \item The statistical distribution of 1,000 km distance absolute magnitude in SDSS $gri$ and clear bands shows that Starlink v1.0 is the brightest mega-constellation bus, followed by Starlink v1.5 and VisorSat. OneWeb satellites are significantly fainter than all versions of Starlink;
  \item The solar phase angle (SPA) is the primary dependent variable of the Starlink satellites' magnitude function, which is well-fitted and can characterize the trend of their brightness; the OneWeb satellites have complex phase-magnitude distributions.
  \item The Starlink satellites featuring the visor design have notably affected brightness mitigation. Based on observation data, VisorSat has a median extinction ratio of 55.1\%, while Starlink v1.5 has a median extinction ratio of 40.4\%. This suggests that Starlink v1.5 utilizes a distinct extinction design compared to VisorSat, leading to significant differences in phase function and multicolor characteristics. While Starlink v1.5's overall extinction ability at small SPA is slightly lower than VisorSat's, the scattered sunlight remains noteworthy.
  \item Satellites of mega-constellations have distinguished color feature. Long-term or multiple measurements of multicolor photometry are helpful for satellite identification.
  \item Based on our observations, implementing an extinction design for mega-constellations can help reduce the impact of scattered light on astronomical observations. Notably, most Starlink Gen1 satellites do not meet the expected brightness mitigation agreement, so it is essential to notice the potential impact of LEO mega-constellations on astronomy.
\end{enumerate}

\section*{Acknowledgements}
 We sincerely appreciate all open access data providers and the constructive comment given by the reviewers. This work was funded by the National Key R\&D Program of China (2022ZD0117401), National Natural Science Foundation of China (NSFC No.12273063), and CAS Key Technology Talent Program.

%%%%%%%%%%%%%%%%%%%%%%%%%%%%%%%%%%%%%%%%%%%%%%%%%%
\section*{Data Availability}

% The inclusion of a Data Availability Statement is a requirement for articles published in MNRAS. Data Availability Statements provide a standardised format for readers to understand the availability of data underlying the research results described in the article. The statement may refer to original data generated in the course of the study or to third-party data analysed in the article. The statement should describe and provide means of access, where possible, by linking to the data or providing the required accession numbers for the relevant databases or DOIs.
The data underlying this article will be shared on reasonable request to the corresponding author.



\bibliographystyle{mnras}
\bibliography{ref_final} % if your bibtex file is called example.bib

%\begin{thebibliography}{99}
%\end{thebibliography}

%%%%%%%%%%%%%%%%%%%%%%%%%%%%%%%%%%%%%%%%%%%%%%%%%%

%%%%%%%%%%%%%%%%% APPENDICES %%%%%%%%%%%%%%%%%%%%%

% \appendix

% \section{Some extra material}

% If you want to present additional material which would interrupt the flow of the main paper,
% it can be placed in an Appendix which appears after the list of references.

%%%%%%%%%%%%%%%%%%%%%%%%%%%%%%%%%%%%%%%%%%%%%%%%%%

% Don't change these lines
\bsp	% typesetting comment
\label{lastpage}
\end{document}